\documentclass[11pt,a4paper]{article}
\input amssymb.sty
\begin{document}
\newcommand{\Si}{\Sigma}
\newcommand{\tr}{{\rm tr}}
\newcommand{\ad}{{\rm ad}}
\newcommand{\Ad}{{\rm Ad}}
\newcommand{\ti}[1]{\tilde{#1}}
\newcommand{\om}{\omega}
\newcommand{\Om}{\Omega}
\newcommand{\de}{\delta}
\newcommand{\al}{\alpha}
\newcommand{\te}{\theta}
\newcommand{\vth}{\vartheta}
\newcommand{\be}{\beta}
\newcommand{\la}{\lambda}
\newcommand{\La}{\Lambda}
\newcommand{\D}{\Delta}
\newcommand{\ve}{\varepsilon}
\newcommand{\ep}{\epsilon}
\newcommand{\vf}{\varphi}
\newcommand{\vfh}{\varphi^\hbar}
\newcommand{\vfe}{\varphi^\eta}
\newcommand{\fh}{\phi^\hbar}
\newcommand{\fe}{\phi^\eta}
\newcommand{\G}{\Gamma}
\newcommand{\ka}{\kappa}
\newcommand{\ip}{\hat{\upsilon}}
\newcommand{\Ip}{\hat{\Upsilon}}
\newcommand{\ga}{\gamma}
\newcommand{\ze}{\zeta}
\newcommand{\si}{\sigma}

\def\clA{\mathcal{A}}
\def\clG{\mathcal{G}}
\def\clR{\mathcal{R}}
\def\clU{\mathcal{U}}
\def\clO{\mathcal{O}}
\def\clL{\mathcal{L}}
\def\clZ{\mathcal{Z}}

\def\bfa{{\bf a}}
\def\bfb{{\bf b}}
\def\bfc{{\bf c}}
\def\bfd{{\bf d}}
\def\bfe{{\bf e}}
\def\bff{{\bf f}}
\def\bfm{{\bf m}}
\def\bfn{{\bf n}}
\def\bfp{{\bf p}}
\def\bfu{{\bf u}}
\def\bfv{{\bf v}}
\def\bft{{\bf t}}
\def\bfx{{\bf x}}
\def\bfg{{\bf g}}
\def\bfC{{\bf C}}
\def\bfS{{\bf S}}
\def\bfJ{{\bf J}}
\def\bfI{{\bf I}}
\def\bfP{{\bf P}}
\def\bfr{{\bf r}}
\def\bfU{{\bf U}}

\def\bfal{\breve{\al}}
\def\bfbe{\breve{\be}}
\def\bfga{\breve{\ga}}
\def\bfnu{\breve{\nu}}
\def\bfsi{\breve{\sigma}}

\def\hS{{\hat{S}}}

\newcommand{\li}{\lim_{n\rightarrow \infty}}
\def\mapright#1{\smash{
\mathop{\longrightarrow}\limits^{#1}}}

\newcommand{\mat}[4]{\left(\begin{array}{cc}{#1}&{#2}\\{#3}&{#4}
\end{array}\right)}
\newcommand{\thmat}[9]{\left(
\begin{array}{ccc}{#1}&{#2}&{#3}\\{#4}&{#5}&{#6}\\
{#7}&{#8}&{#9}
\end{array}\right)}

\newcommand{\thch}[4]{\theta\left[
\begin{array}{c}{#1}\\{#2}
\end{array}
\right]({#3};{#4})}

\newcommand{\beq}[1]{\begin{equation}\label{#1}}
\newcommand{\eq}{\end{equation}}
\newcommand{\beqn}[1]{\begin{eqnarray}\label{#1}}
\newcommand{\eqn}{\end{eqnarray}}
\newcommand{\p}{\partial}
\newcommand{\di}{{\rm diag}}
\newcommand{\oh}{\frac{1}{2}}
\newcommand{\su}{{\bf su_2}}
\newcommand{\uo}{{\bf u_1}}
\newcommand{\SL}{{\rm SL}(2,{\mathbb C})}
\newcommand{\GLN}{{\rm GL}(N,{\mathbb C})}
\newcommand{\PGLN}{{\rm PGL}(N,{\mathbb C})}

\def\sln{{\rm sl}(N, {\mathbb C})}
\def\sl2{{\rm sl}(2, {\mathbb C})}
\def\SLN{{\rm SL}(N, {\mathbb C})}
\def\SLT{{\rm SL}(2, {\mathbb C})}
\newcommand{\gln}{{\rm gl}(N, {\mathbb C})}
\newcommand{\PSL}{{\rm PSL}_2( {\mathbb Z})}
\def\f1#1{\frac{1}{#1}}
\def\lb{\lfloor}
\def\rb{\rfloor}
\def\sn{{\rm sn}}
\def\cn{{\rm cn}}
\def\dn{{\rm dn}}
\newcommand{\rar}{\rightarrow}
\newcommand{\upar}{\uparrow}
\newcommand{\sm}{\setminus}
\newcommand{\ms}{\mapsto}
\newcommand{\bp}{\bar{\partial}}
\newcommand{\bz}{\bar{z}}
\newcommand{\bw}{\bar{w}}
\newcommand{\bA}{\bar{A}}
\newcommand{\bL}{\bar{L}}
\newcommand{\btau}{\bar{\tau}}

\newcommand{\Sh}{\hat{S}}
\newcommand{\vtb}{\theta_{2}}
\newcommand{\vtc}{\theta_{3}}
\newcommand{\vtd}{\theta_{4}}

\def\mC{{\mathbb C}}
\def\mZ{{\mathbb Z}}
\def\mR{{\mathbb R}}
\def\mN{{\mathbb N}}
\def\mP{{\mathbb P}}

\def\frak{\mathfrak}
\def\gg{{\frak g}}
\def\gJ{{\frak J}}
\def\gS{{\frak S}}
\def\gL{{\frak L}}
\def\gG{{\frak G}}
\def\gk{{\frak k}}
\def\gK{{\frak K}}
\def\gl{{\frak l}}
\def\gh{{\frak h}}
\def\gH{{\frak H}}

\newcommand{\ran}{\rangle}
\newcommand{\lan}{\langle}
\def\f1#1{\frac{1}{#1}}
\def\lb{\lfloor}
\def\rb{\rfloor}
\newcommand{\slim}[2]{\sum\limits_{#1}^{#2}}

\newcommand{\sect}[1]{\setcounter{equation}{0}\section{#1}}
\renewcommand{\theequation}{\thesection.\arabic{equation}}
\newtheorem{predl}{Proposition}[section]
\newtheorem{defi}{Definition}[section]
\newtheorem{rem}{Remark}[section]
\newtheorem{cor}{Corollary}[section]
\newtheorem{lem}{Lemma}[section]
\newtheorem{theor}{Theorem}[section]

\begin{center}
{\Large{\bf Integrable systems associated with generalized Sklyanin algebra}}\\

\vspace{5mm} Yu.Chernyakov, \footnote{e-mail: chernyakov@itep.ru}

\vspace{3mm} {\it
Institute for Theoretical and Experimental Physics, Moscow.\\
}

\vspace{5mm}
\end{center}

\begin{abstract}
Using the point fusion procedure we obtain the new integrable
systems from the Elliptic Schlesinger system (ESS). These new
systems have the pole orders higher than one in the matrix of the
Lax operator. Quadratic Poisson algebras on the phase space of the
new systems generalize the Sklyanin algebras and have the
graduated structure.
\end{abstract}
\today

\section{Introduction}
The examples of the systems with Lax operators having the pole
orders higher than one were considered in papers \cite{Ch} and
\cite{MPR}. These systems were obtained by point fusion procedure
called in algebra Inonu-Wigner (\cite{IW}) contraction. The main
idea of this method consists of finding such decomposition of the
variables, which gives us the pole order in the matrix of Lax
operator being higher than one at some marked point. It implies
the existence of Hamiltonian and Casimir numbers in this new
system being the same as in the initial one. The main object of
this procedure in the present paper is the Lax operator of the
Elliptic Schlesinger system (ESS) which was considered in
\cite{ChLOZ}.

Historically ESS appeared as an approach to the decision of
Riman's problem about searching the differential equations with
regular singularities and intended monodromy data: $M_{j}, \
j=1,...n \ (\Psi \rightarrow \Psi M_{j})$, where $\Psi$ is the
solution of initial linear system. In \cite{Sch} L.Schlesinger had
considered the first order system of differential equation on
$\mC\mP^1$

$$ \left\{
 \begin{array}{l}
\left(\p_z+\sum_{j=1}^{n}\frac{\bfS^j}{z-x_j}\right)\Psi=0\,\\
  \p_{\bw}\Psi=0\,
  \end{array}
 \right.
$$

and found the preserving conditions on matrix $S^{j}$ with respect
to the changing the point positions $x_{j}$. He obtained the
system of first order differential equations for $n>3$ matrices
$\bfS^j$ $(j=1,\ldots,n)$, depending on $n$ points
$x_k\in\mC\mP^1$ \beq{S0}
\p_k\bfS^j=\frac{[\bfS^k,\bfS^j]}{x_k-x_j}\,,~(k\neq j)\,,
~~\p_k=\p_{x_k}\,, \eq \beq{S} \p_k\bfS^k=-\sum_{j\ne
k}\frac{[\bfS^k,\bfS^j]}{x_k-x_j}\,. \eq This system is
non-autonomous Hamiltonian system and has the Hamiltonian form
with respect to the linear (Lie-Poisson) brackets on $\sln$. The
Hamiltonian
$$
H_k=\sum_{j\neq
k}\frac{\lan\bfS^k\bfS^j\ran}{x_k-x_j}\,~~(\lan\,\,\ran=\tr)
$$
defines the evolution with respect to the time $x_k$.

For two by two matrices and four marked points the Schlesinger
system is equivalent to the Painlev{\'e} VI equation
(\cite{LOZ},\cite{ChLOZ}). In this case the position of three
points can be fixed as $(0,1,\infty)$ while $x_4$ play the role of
an independent variable.

If we replace $\mC\mP^1$ by an elliptic curve, we define a similar
system (the elliptic Schlesinger system (ESS)). In this case, in
addition to the coordinates of the marked points, a new
independent variable appears inevitably. It is the modular
parameter of the curve, and thereby we have an additional new
Hamiltonian. This system was introduced originally by Takasaki
\cite{Ta}. His derivation is based on the quasi-classical limit of
the quantum $SU(N)$ version of the XYZ model. In \cite{ChLOZ} was
obtained as symplectic quotient of the symplectic space of
connections of principle bundles of degree one over the elliptic
curves with $n$ marked points. This approach was previously
developed in \cite{LO}.

Let us give a short description of ESS.

\paragraph{Elliptic Schlesinger System.}

Let us consider elliptic curve $\Si_\tau=\mC/(\mZ+\tau\mZ)$ with
the modular parameter $\tau$, $( \Im (\tau)>0)$ and
$$
D_n=(x_1,\ldots,x_n)\,,~x_j\neq x_k\,,~x_k\in\Si_\tau
$$
be the divisor  of non-coincident  points with the condition
\beq{d} \sum x_j\in(\mZ+\tau\mZ)\,. \eq

Consider the space ${\cal P}_{n,N}$ of $n$ copies of the Lie
coalgebra $\gg^*\sim\sln^*$, related to the points of the divisor.
\beq{lb} {\cal P}_{n,N}=\oplus_{j=1}^n\gg_j^*\,,~~
\gg_j^*=\{\bfS^j=\sum_{\al\in\ti{\mZ}^{(2)}_N} S_\al^j T^\al \}\,,
\eq where $T^\al$ is the basis element of $GL(N,\mathbb{C})$ (see
Appendix B).

Introduce operators acting from ${\cal P}_{n,N}$ to the dual
$\oplus_{j=1}^n\gg_j$ \beq{phi} \bfI_{kj}\,:\,\gg_k^*\to\gg_j\,,~~
S^k_\ga   \mapsto(I_{kj})_\ga S^j_\ga\,,~
  (I_{kj})_\ga=\varphi_{\ga}(x_j-x_k) \,,
\eq \beq{J} \bfJ_{jj}\,:\,\gg_j^*\to\gg_j\,,~~ S^j_\ga\mapsto
J_\ga S^j_\ga\,,~J_\ga=E_2(\bfga)\,, \eq \beq{j}
\bfJ_{kj}\,:\,\gg_k^*\to\gg_j\,,~~ S^k_\ga\mapsto (J_{kj})_\ga
S^j_\ga\,,~(J_{kj})_\ga=f_\ga(x_j-x_k) \eq where $\varphi_\ga(x)$,
$E_2(\bfga)$, $f_\ga(x)$ are defined in Appendix B.

The positions of the marked points $x_j\in D_n$ and the modular
parameter $\tau$ are local coordinates in an open cell in the
moduli space ${\cal M}_{1,n}$ of elliptic curves with $n$ marked
points. They play the role of times.

\begin{defi}.
The elliptic Schlesinger system (ESS) is the consistent dynamical
system on ${\cal P}_{n,N}$ with independent variables from ${\cal
M}_{1,n}$ \beq{1} \p_j\bfS^k=[\bfI_{kj}(\bfS^j),\bfS^k]\,,~(k\neq
j)\,, ~~\p_k=\p_{x_k}\,, \eq \beq{2} \p_k\bfS^k=-\sum_{j\ne
k}[\bfI_{jk}(\bfS^j),\bfS^k]\,, \eq \beq{3} \p_\tau\bfS^j=
\sum_{k\neq j}\f1{2\pi\imath}[\bfS^j,\bfJ_{kj}(\bfS^k)]
+\f1{4\pi\imath}[\bfS^j,\bfJ_{jj}(\bfS^j)]\,, \eq where the
commutators are understand as the coadjoint action of $\gg_j$ on
$\gg^*_j$.
\end{defi}

In the basis $t^\al\,$ $(\al\in\ti{\mZ}^{(2)}_N)$ (\ref{db}) the
ESS takes the form \beq{2.1} \p_kS_\al^j=
\sum_{\ga\in\ti{\mZ}^{(2)}_N)}\bfC(\ga,\al)S_\ga^k
S^j_{\al-\ga}\varphi_{\ga}(x_j-x_k)\,,~~(k\neq j)\,, \eq
\beq{2.1a}
 \p_kS_\al^k=\sum_{\ga\in\ti{\mZ}^{(2)}_N)}\bfC(\ga,\al)
 \sum_{j\neq k}S_{\al-\ga}^j S^k_{\ga}\varphi_{\al-\ga}(x_k-x_j)\,,
\eq \beq{2.2} \p_\tau
S^k=\f1{2\pi\imath}\sum_{\ga\in\ti{\mZ}^{(2)}_N)}\bfC(\al,\ga)
\left( \sum_{k\neq j} S_{\al-\ga}^kS^j_{\ga}f_\ga(x_k-x_j)
+S_\ga^kS_{-\ga}^kE_2(\bfga) \right)\,. \eq

\begin{rem}.
In the  rational limit ($\Im m\tau\to\infty$) (\ref{2.1}) and
(\ref{2.1a}) pass to the standard Schlesinger system (\ref{S0}),
(\ref{S}) (see (\ref{A.3a})).
\end{rem}

As in the rational case the ESS has some fundamental properties
  The space ${\cal P}_{n,N}^{1}$ is Poisson with respect to the
linear Lie-Poisson brackets on $\gg^*$ \beq{lpb}
\{S_\al^j,S_\be^k\}_1=\de^{jk}\bfC(\al,\be)S^j_{\al+\be} \eq ESS
is a non-autonomous Hamiltonian system on ${\cal P}_{n,N}$
  \beq{2.3}
  \p_k\bfS^j=\{H_{k},\bfS^j,\}_1\,,~~\p_k=\p_{x_k}\,,~(1,\ldots,n)\,,
  \eq
  \beq{2.4}
  \p_\tau \bfS^j=\{H_{0},\bfS^j\}_1\,,
\eq where \beq{2.5} H_{k}=-\sum_{j\neq
k}\lan\bfI_{kj}(\bfS^k)\bfS^j)\ran=
   - \sum_{j\neq k}\sum_{\ga\in\ti{\mZ}^{(2)}_N}
S_\ga^kS_{-\ga}^j\varphi_\ga(x_j-x_k)\,, \eq \beq{2.6}
H_\tau=H_0=-\f1{2\pi\imath} \left( \sum_{k\neq j}\lan \bfS^j
\bfJ_{kj}(\bfS^k)\ran+ \sum_j\lan \bfS^j \bfJ_{jj}(\bfS^j)\ran
\right) \eq
$$
=-\f1{2\pi\imath} \left(\sum_{k\neq
j}\sum_{\ga\in\ti{\mZ}^{(2)}_N}
S_\ga^jS_{-\ga}^kf_\ga(x_k-x_j)+\sum_{
j}\sum_{\ga\in\ti{\mZ}^{(2)}_N} S_\ga^jS_{-\ga}^jE_2(\bfga)
\right)\,.
$$
The brackets (\ref{lpb}) are degenerate. The symplectic leaves are
$n$ copies of coadjoint orbits ${\cal O}_j$ $(j=1,\ldots,n)$ of
$\SLN$. Assume that all orbits are generic, and let $c^\mu(j)$ be
corresponding  Casimir functions of order $\mu$
$(\mu=2,\ldots,N)$. The phase space of ESS is \beq{ps1} {\cal
R}_{n,N} \sim {\cal P}_{n,N} /\{c^\mu(j)=c^{\mu}(j)_{0}\}
\sim\prod{\cal O}_j\,, \eq \beq{dps1} \dim{\cal
R}_{n,N}=nN(N-1)\,. \eq The ESS can be considered as a system of
interacting non-autonomous $\SLN$ Euler-Arnold tops, where
operators (\ref{phi}), (\ref{J}), (\ref{j}) play the role of the
inverse inertia tensors.

\section{Classical integrable systems obtained from ESS}
\paragraph{ESS in the case of two marked point and $N=2$.}
The Lax operator of ESS in the case of $n$ marked points has the
following form:

\beq{cl00} L(z)=-\f1{N} E_1(z)T_0+
\sum_{j=1}^n\sum_{\al\in\ti{\mZ}^{(2)}_N}S_\al^j\varphi_\al(z-x_j)T_\ga\,.
\eq

where $T_{0}$ and $T_{\alpha}$ are basis elements of
$GL(N,\mathbb{C})$.

Let us consider the modified Lax operator (\cite{ChLOZ})

\beq{cl01} L^{group}(z) = S_{0} T_{0}  + \sum_{j}^{n} \left(
S_{0}^{j}E_{1}(z-x_{j}) T_{0} + \sum_{\alpha} S_{\alpha}^{j}
\varphi_{\alpha}(z-x_{j})  \right) T_{\alpha} \,, \eq where we
attribute to the marked points of the divisor $D_n$ $n$ copies of
the
 $\GLN$-valued elements
$$
x_j\to S_0^jT_0+\bfS^j=\sum_{a\in\mZ^{(2)}_N} S_a^jT_a\,,
$$
adding to this set a variable $S_0\in\mC$. So, it defines
$$
{\cal P}_{n,N}^{+}=\{S_0\,,\,(S_0^j\,,
\bfS^j\,,j=1,\ldots,n)\,|\,\sum_{j=1}^nS^j_0=0\}\,.
$$

It is possible to obtain the equation of motion for ESS from the
quadratic brackets on the space ${\cal P}_{n,N}^{+}$, extracting
them from the classical exchange algebra \beq{cl13}
\left\{L^{group}(z),L^{group}(w)\right\}= [r(z-w),L^{group}(z)
\otimes L^{group}(w)]\,. \eq

where $r$ is the classical Belavin-Drinfeld r-matrix.

Let us consider the first occurrence: $N=2$ and $n=2$ is a number
of the marked points. Then the Lax operator takes the following
form:

 $$ L^{group}(z)= \left( S_{0} + S_{0}^{a}E_{1}(z-x_{a}) +
S_{0}^{b}E_{1}(z-x_{b}) \right) \sigma_{0} +$$
\beq{cl11}
+ \sum_{\alpha} \left(
S_{\alpha}^{a} \varphi_{\alpha}(z-x_{a}) + S_{\alpha}^{b}
\varphi_{\alpha}(z-x_{b}) \right) \sigma_{\alpha} \,, \eq

$$S_{0}^{a} + S_{0}^{b} = 0.$$

Note, that for $N=2$ the basis $T_{\al}$ is proportional to the
basis of the Pauli matrices. The dimension of the phase space
$R_{2,2}$ is four. The Hamiltonians of this system are $S_{0}$ and
$S_{0}^{a}$. The quadratic brackets:

\beq{cl12a}
\partial_{\tau}S_{\alpha}^a = \frac{1}{2}\left\{S_{\alpha}^a,
S_0\right\} =
\eq

$$= i\varepsilon_{\alpha\beta\gamma} \left(E_2(\ga) -
E_2(\be)\right) S_\be^a S_\ga^a - i\varepsilon_{\alpha\beta\gamma}
\varphi_{\ga}^{'}(x_{ab}) S_\be^a S_\ga^b +
i\varepsilon_{\alpha\beta\gamma} \varphi_{\be}^{'}(x_{ab}) S_\be^b
S_\ga^a \,,$$

\beq{cl12a}
\partial_{x_a}S_{\alpha}^a = -\partial_{x_b}S_{\alpha}^a =
 \frac{1}{2}\left\{S_\al^a, S_0^a\right\} =
\eq

$$= i\varepsilon_{\alpha\beta\gamma} \varphi_{\be}(x_{ab}) S_\be^b
S_\ga^a - i\varepsilon_{\alpha\beta\gamma} \varphi_{\ga}(x_{ab})
S_\ga^b S_\be^a \,. \,$$

\paragraph{System obtained via two point fusion.}
Let us fulfil the following coordinate transformation and
decomposition of the variables $S$ (\cite{Ch}):

\beq{cl1} x_{b}=x_{a} + \varepsilon, \ \
S_{0}^{a}=c_{a,0}^{0}S_{0}^{0} +
c_{a,1}^{1}S_{0}^{1}\varepsilon^{-1}, \ \
S_{\alpha}^{a}=c_{a,\alpha}^{0}S_{0}^{0} +
c_{a,\alpha}^{1}S_{0}^{1}\varepsilon^{-1} \,, \eq where $c$ are some coefficients.
 Taking the limit $\varepsilon
\rightarrow 0$ and putting some additional conditions on
coefficients (the requirements of the absence of singularities) we
get the new Lax operator $L_{fusion}(z)$
\beq{cl2} L_{fusion}(z) = \left( S_{0} + S_{0}^{1}E_{2}(z) \right)
\sigma_{0} + \sum_{\alpha} \left( S_{\alpha}^{0}
\varphi_{\alpha}(z) + S_{\alpha}^{1} \varphi_{\alpha}^{'}(z)
\right) \sigma_{\alpha} \,, \eq
where we put the position $x_{a}$ equal to $0$. The dimension of
the phase space
$$ \mathcal{P}_{f,2} = \left( S_{0},S_{0}^{1}, S_{\alpha}^{0}, S_{\alpha}^{1} \right) $$
is 8 and it is equal to the dimension of ${\cal P}_{n,N}^{+}$
before point fusion.

To find the exchange relations between the new variables $S$ we
consider the equation (\ref{cl13}) and the new Lax operator
(\ref{cl2}). These exchange relations are the coefficients at the
function products  $1,\ E_{2}(z),\ \varphi_{\alpha}(z),\
\varphi_{\alpha}^{'}(z)$ and $1,\ E_{2}(w),\ \varphi_{\alpha}(w),\
\varphi_{\alpha}^{'}(w)$. So we get
\begin{predl}. The space  $\mathcal{P}_{f_{2},2}$ is Poisson with respect to the
quadratic brackets
\end{predl}

\beq{cl3} \left\{S_0,S_0^1\right\} = 0 \,, \eq

\beq{cl4}
\begin{array}{c}
\left\{S_{\alpha}^0, S_{\beta}^0\right\} =
2i\varepsilon_{\alpha\beta\gamma} S_\be^0 S_\ga^0 -
2i\varepsilon_{\alpha\beta\gamma} E_2(\ga) S_0^1 S_\ga^0 \,,\\

\left\{S_\al^1, S_\be^0\right\} =
2i\varepsilon_{\alpha\beta\gamma} S_0 S_\ga^1 -
2i\varepsilon_{\alpha\beta\gamma} E_2(\al) S_0^1
S_\ga^1 \,,\\

\\

\left\{S_\al^1, S_\be^1\right\} =
2i\varepsilon_{\alpha\beta\gamma} S_0^1 S_\ga^0 \,,
\end{array}
 \eq

\beq{cl5}
\begin{array}{c}
\left\{S_{\alpha}^0, S_0\right\} =
2i\varepsilon_{\alpha\beta\gamma} \left(E_2(\ga) - E_2(\be)\right)
S_0 S_\ga^0 + 2i\varepsilon_{\alpha\beta\gamma}  E_2(\al)
\left(E_2(\ga) - E_2(\be)\right) S_\be^1
S_\ga^1\,,\\

\\

\left\{S_\al^1, S_0\right\} = 2i\varepsilon_{\alpha\beta\gamma}
E_2(\be) S_\be^0 S_\ga^1 - 2i\varepsilon_{\alpha\beta\gamma}
E_2(\ga) S_\be^1
S_\ga^0 \,,\\

\\

\left\{S_\al^0, S_0^1\right\} = 2i\varepsilon_{\alpha\beta\gamma}
\left(E_2(\ga) - E_2(\be)\right)
 S_\be^1
S_\ga^1 \,,\\

\left\{S_\al^1, S_0^1\right\} = -
2i\varepsilon_{\alpha\beta\gamma}
 S_\be^0 S_\ga^1 + 2i\varepsilon_{\alpha\beta\gamma}
 S_\be^1 S_\ga^0 \,.
\end{array}
 \eq

{\it Proof}:

Let us consider the classical Poisson brackets \beq{ad31}
\{L(z),L(w)\}= [r(z-w),L(z) \otimes L(w)]\,, \eq where the Lax
operator has the following form: \beq{ad32} L(z) = L_{fusion}(z) =
\left( S_{0} + S_{0}^{1}E_{2}(z) \right) \sigma_{0} +
\sum_{\alpha} \left( S_{\alpha}^{0} \varphi_{\alpha}(z) +
S_{\alpha}^{1} \varphi_{\alpha}^{'}(z) \right) \sigma_{\alpha} \,,
\eq

In the l.h.s. for the matrix element
$\sigma_{\alpha}\otimes\sigma_{\be}$ in (\ref{ad31}) we have the
sum consisted of the following terms:
$$\{S_{\alpha}^{0},S_{\beta}^{0}\}\varphi_{\alpha}(z)\varphi_{\beta}(w),
\
\{S_{\alpha}^{0},S_{\beta}^{1}\}\varphi_{\alpha}(z)\varphi_{\beta}^{'}(w),$$
$$
\{S_{\alpha}^{1},S_{\beta}^{0}\}\varphi_{\alpha}^{'}(z)\varphi_{\beta}(w),
\
\{S_{\alpha}^{1},S_{\beta}^{1}\}\varphi_{\alpha}^{'}(z)\varphi_{\beta}^{'}(w),$$

and for the matrix element $\sigma_{\alpha} \otimes I$:

$$\{S_{\alpha}^{0},S_{0}\}\varphi_{\alpha}(z), \
\{S_{\alpha}^{0},S_{0}^{1}\}\varphi_{\alpha}(z)E_2(w), \
\{S_{\alpha}^{1},S_{0}\}\varphi_{\alpha}^{'}(z), \
\{S_{\alpha}^{1},S_{0}^{1}\}\varphi_{\alpha}^{'}(z)E_2(w).$$

In the r.h.s. we have for the same matrix elements the sums
consisted of the following terms:
$$2i\varepsilon_{\alpha\beta\gamma}S_{\ga}^{0}S_{0}
\left(\varphi_{\alpha}(z-w)\varphi_{\ga}(w) -
\varphi_{\be}(z-w)\varphi_{\ga}(z)\right),$$
$$2i\varepsilon_{\alpha\beta\gamma}S_{\ga}^{0}S_{0}^{1}
\left(\varphi_{\alpha}(z-w)\varphi_{\ga}(w)E_2(z) -
\varphi_{\be}(z-w)\varphi_{\ga}(z)E_2(w)\right),$$
$$2i\varepsilon_{\alpha\beta\gamma}S_{\ga}^{1}S_{0}
\left(\varphi_{\alpha}(z-w)\varphi_{\ga}^{'}(w) -
\varphi_{\be}(z-w)\varphi_{\ga}^{'}(z)\right),$$
$$2i\varepsilon_{\alpha\beta\gamma}S_{\ga}^{1}S_{0}^{1}
\left(\varphi_{\alpha}(z-w)\varphi_{\ga}^{'}(w)E_2(z) -
\varphi_{\be}(z-w)\varphi_{\ga}^{'}(z)E_2(w)\right)$$

and

$$2i\varepsilon_{\alpha\beta\gamma}S_{\ga}^{0}S_{\be}^0
\left(\varphi_{\ga}(z-w)\varphi_{\be}(z)\varphi_{\ga}(w) -
\varphi_{\be}(z-w)\varphi_{\ga}(z)\varphi_{\be}(w)\right),$$
$$2i\varepsilon_{\alpha\beta\gamma}S_{\ga}^{1}S_{\be}^0
\left(\varphi_{\ga}(z-w)\varphi_{\be}(z)\varphi_{\ga}^{'}(w) -
\varphi_{\be}(z-w)\varphi_{\ga}^{'}(z)\varphi_{\be}(w)\right),$$
$$2i\varepsilon_{\alpha\beta\gamma}S_{\ga}^{0}S_{\be}^1
\left(\varphi_{\ga}(z-w)\varphi_{\be}^{'}(z)\varphi_{\ga}(w) -
\varphi_{\be}(z-w)\varphi_{\ga}(z)\varphi_{\be}^{'}(w)\right),$$
$$2i\varepsilon_{\alpha\beta\gamma}S_{\ga}^{1}S_{\be}^1
\left(\varphi_{\ga}(z-w)\varphi_{\be}^{'}(z)\varphi_{\ga}^{'}(w) -
\varphi_{\be}(z-w)\varphi_{\ga}^{'}(z)\varphi_{\be}^{'}(w)\right)$$

As a result of the expansion in function product for the matrix
element $\sigma_{\alpha}\otimes\sigma_{\be}$ we get:

$$2i\varepsilon_{\alpha\beta\gamma}S_{\ga}^{0}S_{0}
\left(\varphi_{\alpha}(z-w)\varphi_{\ga}(w) -
\varphi_{\be}(z-w)\varphi_{\ga}(z)\right) =
2i\varepsilon_{\alpha\beta\gamma}
\varphi_{\alpha}(z)\varphi_{\be}(w)S_{\ga}^{0}S_{0},$$

$$2i\varepsilon_{\alpha\beta\gamma}S_{\ga}^{0}S_{0}^{1}
\left(\varphi_{\alpha}(z-w)\varphi_{\ga}(w)E_2(z) -
\varphi_{\be}(z-w)\varphi_{\ga}(z)E_2(w)\right) =$$
$$=2i\varepsilon_{\alpha\beta\gamma}
\left(\varphi_{\alpha}(z)\varphi_{\be}(w)E_2(\ga) -
\varphi_{\alpha}^{'}(z)\varphi_{\be}^{'}(w)\right)S_{\ga}^{0}S_{0}^{1},$$

$$2i\varepsilon_{\alpha\beta\gamma}S_{\ga}^{1}S_{0}
\left(\varphi_{\alpha}(z-w)\varphi_{\ga}^{'}(w) -
\varphi_{\be}(z-w)\varphi_{\ga}^{'}(z)\right)=$$
$$=2i\varepsilon_{\alpha\beta\gamma}
\left(\varphi_{\alpha}^{'}(z)\varphi_{\be}(w) +
\varphi_{\alpha}(z)\varphi_{\be}^{'}(w)\right)S_{\ga}^{1}S_{0} ,$$

$$2i\varepsilon_{\alpha\beta\gamma}S_{\ga}^{1}S_{0}^{1}
\left(\varphi_{\alpha}(z-w)\varphi_{\ga}^{'}(w)E_2(z) -
\varphi_{\be}(z-w)\varphi_{\ga}^{'}(z)E_2(w)\right)=$$
$$=
2i\varepsilon_{\alpha\beta\gamma}
\left(\varphi_{\alpha}^{'}(z)\varphi_{\be}(w)E_2(\be) +
\varphi_{\alpha}(z)\varphi_{\be}^{'}(w)E_2(\al)
\right)S_{\ga}^{1}S_{0}^{1}.
$$

and for the matrix element $\sigma_{\alpha} \otimes I$:

$$2i\varepsilon_{\alpha\beta\gamma}S_{\ga}^{0}S_{\be}^0
\left(\varphi_{\ga}(z-w)\varphi_{\be}(z)\varphi_{\ga}(w) -
\varphi_{\be}(z-w)\varphi_{\ga}(z)\varphi_{\be}(w)\right)=$$
$$=2i\varepsilon_{\alpha\beta\gamma}
\varphi_{\al}(z)\left(E_2(\ga)-E_2(\be)\right)
S_{\ga}^{0}S_{\be}^0,$$

$$2i\varepsilon_{\alpha\beta\gamma}S_{\ga}^{1}S_{\be}^0
\left(\varphi_{\ga}(z-w)\varphi_{\be}(z)\varphi_{\ga}^{'}(w) -
\varphi_{\be}(z-w)\varphi_{\ga}^{'}(z)\varphi_{\be}(w)\right)=$$
$$=2i\varepsilon_{\alpha\beta\gamma}
\left(-\varphi_{\al}^{'}(z)E_2(\be) +
\varphi_{\al}^{'}(z)E_2(w)\right)
 S_{\ga}^{1}S_{\be}^0,$$

$$2i\varepsilon_{\alpha\beta\gamma}S_{\ga}^{0}S_{\be}^1
\left(\varphi_{\ga}(z-w)\varphi_{\be}^{'}(z)\varphi_{\ga}(w) -
\varphi_{\be}(z-w)\varphi_{\ga}(z)\varphi_{\be}^{'}(w)\right)=$$
$$=2i\varepsilon_{\alpha\beta\gamma}
\left(\varphi_{\al}^{'}(z)E_2(\ga) -
\varphi_{\al}^{'}(z)E_2(w)\right)
 S_{\ga}^{0}S_{\be}^1,$$

$$2i\varepsilon_{\alpha\beta\gamma}S_{\ga}^{1}S_{\be}^1
\left(\varphi_{\ga}(z-w)\varphi_{\be}^{'}(z)\varphi_{\ga}^{'}(w) -
\varphi_{\be}(z-w)\varphi_{\ga}^{'}(z)\varphi_{\be}^{'}(w)\right)=$$
$$=2i\varepsilon_{\alpha\beta\gamma}
\left(-\varphi_{\al}(z)E_2(\al)\left(E_2(\be)-E_2(\ga)\right)
+
\varphi_{\al}(z)E_2(w)\left(E_2(\be)-E_2(\ga)\right)\right)
 S_{\ga}^{1}S_{\be}^1.$$\\

After regrouping the terms we get (\ref{cl3}),(\ref{cl4}) and
(\ref{cl5}) as the coefficients at the function products $1,\
E_2(z),\ \varphi_{\alpha}(z),\ \varphi_{\alpha}^{'}(z)$ and $1,\
E_2(w),\ \varphi_{\alpha}(w),\ \varphi_{\alpha}^{'}(w)$. $\Box$\\

Let us write these exchange relations in the form of the tables (1) and (2):\\
\ \\

\begin{center}
Table 1. \ Exchange relations $S_{\alpha},\ S_{\beta}$
\end{center}
$$
\begin{tabular}{|c|c|c|c|c|}
\hline \cline{1-0}
$$ & $$ & $$ & $$ & $$\\

$  \ \ \ \ \ \ \ \ \ \ \ \ 2i\varepsilon_{\alpha\beta\gamma} \cdot$ & $S_0S_\ga^{0}$ & $S_0S_\ga^{1}$ & $S_0^{1}S_\ga^{0}$ & $S_0^{1}S_\ga^{1}$ \\
\hline \cline{1-0}
$$ & $$ & $$ & $$ & $$\\

$\left\{S_{\alpha}^0, S_{\beta}^0\right\}$ & $+1$ & $0$ & $-J_\ga$ & $0$ \\
\hline \cline{1-0}
$$ & $$ & $$ & $$ & $$\\

$\left\{S_{\alpha}^1, S_{\beta}^0\right\}$ & $0$ & $+1$ & $0$ & $-J_\al$ \\
\hline \cline{1-0}
$$ & $$ & $$ & $$ & $$\\

$\left\{S_{\alpha}^0, S_{\beta}^1\right\}$ & $0$ & $+1$ & $0$ & $-J_\be$ \\
\hline \cline{1-0}
$$ & $$ & $$ & $$ & $$\\

$\left\{S_{\alpha}^1, S_{\beta}^1\right\}$ & $0$ & $0$ & $+1$ & $0$ \\
\hline
\end{tabular}
$$

and


\begin{center}
Table 2. \ Exchange relations $S_{\alpha},\ S_{0}$
\end{center}

$$
\begin{tabular}{|c|c|c|c|c|}
\hline\cline{1-0}
$$ & $$ & $$ & $$ & $$\\

$ \ \ \ \ \ \ \ \ \ \ \ \ 2i\varepsilon_{\alpha\beta\gamma} \cdot$ & $S_\be^{0} S_\ga^{0}$ & $S_\be^{0}S_\ga^{1}$ & $S_\be^{1}S_\ga^{0}$ & $S_\be^{1}S_\ga^{1}$ \\
\hline\cline{1-0}
$$ & $$ & $$ & $$ & $$\\

$\left\{S_{\alpha}^0, S_0\right\}$ & $+J_{\ga\be}$ & $0$ & $0$ & $+J_\al J_{\ga\be}$\\
\hline\cline{1-0}
$$ & $$ & $$ & $$ & $$\\

$\left\{S_{\alpha}^1, S_0\right\}$ & $0$ & $-J_\be$ & $+J_\ga$ & $0$ \\
\hline\cline{1-0}
$$ & $$ & $$ & $$ & $$\\

$\left\{S_{\alpha}^0, S_0^1\right\}$ & $0$ & $0$ & $0$ & $+J_{\ga\be}$ \\
\hline\cline{1-0}
$$ & $$ & $$ & $$ & $$\\

$\left\{S_{\alpha}^1, S_0^1\right\}$ & $0$ & $-1$ & $+1$ & $0$ \\
\hline
\end{tabular}
$$\\

where $J_\al=E_2(\al), \ \ \ J_{\ga\be} = E_{2}(\ga)-E_{2}(\be)$.
The Jacobi identity for $\mathcal{P}_{f,2}$ follows from the
classical Yang-Baxter equation for $r$ matrix. Note that the
Poisson algebra $\mathcal{P}_{f,2}$ come to the Sklyanin algebra
(\cite{Scl}), if we put $S_{0}^{1}=0, \ S_{\alpha}^{1}=0$.

The quadratic brackets are not degenerate on the orbits. To
describe the system we define the Casimir functions and
Hamiltonians. The equation for the spectral curve has the
following form:

$$
det (L(z) - \lambda I) = 0,$$
$$
 \lambda^2 - Tr(L(z))\lambda +
detL(z)=0,
$$

where $det$ is determinant. The coefficients $TrL(z)$ and
$detL(z)$ of this equation define the Casimir functions and
Hamiltonians of the system. $TrL(z)$ and $detL(z)$ are doubly
periodic functions and they can be decomposed into the basis of
Eisenstien functions.

$$
 \frac{1}{2} TrL(z) = S_{0} + E_2(z) S_{0}^{1},
$$

$$
detL(z) = C_0 + C_1 E_1(z) + C_2 E_2(z) + C_3 E_2^{'}(z) + C_4
E_2^{''}(z).
$$

So, we have two Hamiltonians $S_0$, $S_0^1$ and four the Casimir
functions ($C_1=0$):

$$
C_0 = S_0^2 - 4\eta_1^2(S_0^1)^2 + E_2(\al)(S_\al^0)^2 +$$
$$+\left((E_2(\al))^2 - (E_1(\al))^2E_2(\al) + E_1(\al)E_2^{'}(\al) +
\frac{1}{3} E_2^{''}(\al)\right)(S_\al^1)^2,
$$

$$
C_2 =  S_0S_0^1 + 4\eta_1(S_0^1)^2 - (S_\al^0)^2 +
\left((E_1(\al))^2 - E_2(\al)\right) (S_\al^1)^2,
$$

$$
C_3 = -S_\al S_\al^1,
$$

$$
C_4 = \frac{1}{6}\left((S_0^1)^2 - (S_\al^1)^2\right).
$$

To calculate the Casimir functions we use (\ref{A.2b}) -
(\ref{A.5b}) of Appendix A. The dimension of the phase space
$\mathcal{R}_{f,2}$ (symplectic leaves) is equal to 4.
$$
\mathcal{R}_{f,2} \sim \mathcal{P}_{f,2} / \{ C_i,
i=\overline{1,4} \}
$$

In terms of the quadratic brackets the equations of motion have
the following form:

$$\partial_{t_{0}}S_{\alpha}^0 = \frac{1}{2}\{S_{\alpha}^0, S_0\} =$$

\beq{cl6}
\begin{array}{c}

= i\varepsilon_{\alpha\beta\gamma}
\left(E_2(\ga) - E_2(\be)\right) S_\be^0 S_\ga^0 +
i\varepsilon_{\alpha\beta\gamma} E_2(\al) \left(E_2(\ga) -
E_2(\be)\right) S_\be^1
S_\ga^1\,,\\

\\

\partial_{t_{0}}S_{\alpha}^1 = \frac{1}{2}\{S_\al^1, S_0\} = -i\varepsilon_{\alpha\beta\gamma}
E_2(\be)S_\be^0 S_\ga^1 + i\varepsilon_{\alpha\beta\gamma}
E_2(\ga) S_\be^1
S_\ga^0 \,,\\

\\

\partial_{t_{1}}S_{\alpha}^0 = \frac{1}{2}\{S_\al^0, S_0^1\} =
i\varepsilon_{\alpha\beta\gamma} \left(E_2(\ga) - E_2(\be)\right)
S_\be^1
S_\ga^1 \,,\\

\\

\partial_{t_{1}}S_{\alpha}^1 = \frac{1}{2}\{S_\al^1, S_0^1\} = - i\varepsilon_{\alpha\beta\gamma}
 S_\be^0 S_\ga^1 + i\varepsilon_{\alpha\beta\gamma}
 S_\be^1 S_\ga^0 \,.
\end{array}
 \, \eq


\paragraph{Graduation and the systems via three point fusion.}

Let us do the following observation. One can see from the tables
(3) and (4) that coefficients at $SS$ products have the certain
dependence of the same products. Consider the bracket
 $\left\{S_{\alpha}^0, S_{\beta}^0\right\}$, we note that
$S_0S_\ga^{0}$ differs from $S_0^{1}S_\ga^{0}$ by one of the
multipliers. The coefficients take the values $+1$ and $-J_\ga$
correspondingly. For the bracket $\left\{S_{\alpha}^0,
S_{\beta}^1\right\}$ we have the same situation: the product
$S_0S_\ga^{1}$ differs from $S_0^{1}S_\ga^{1}$ one by the same
multipliers $S_0$ and $S_0^{1}$. The coefficients take the values
$+1$ and $-J_\be$. Now the coefficient at $S_0^{1}S_\ga^{1}$
depend on $\be$. Taking into account
$J_u=E_2(u)=-\frac{\partial}{\partial u} E_1(u)$, we see that each
function $J$ has the degree of the operator
$\frac{\partial}{\partial u}$ proportional to 1, but it takes the
different value. It is possible to say that the function $+1$ has
the degree of the operator $\frac{\partial}{\partial u}$ equal to
0. It is possible to see the regularization in the coefficient
positions for the other brackets. So, we can define the notion of
the graduation in the following manner. We consider the Lax
operator $L_{fusion}(z)$ (\ref{cl2}) and assign that the
graduation of variables $S$ - $grad(S) \equiv [S]$ is proportional
to the modulus of the pole order of the function at $S$ in
(\ref{cl2}). It is possible to assume that in the expression of
the each Poisson bracket all the terms have the equal graduation.
Here we take into account that the graduation of the coefficients
(which is proportional to the degree of the operator) is opposite
in sign to the $SS$ products. The graduation of each term is the
sum of the graduations:
 $$[J_\be S_0^{1}S_\ga^{1}] = [J_\be]
+ [S_0^{1}] + [S_\ga^{1}].$$ From this approach the interesting
fact is that the operation of taking Poisson bracket acquire the
graduation too. $$[\left\{S_{\alpha}^0,
S_{\beta}^0\right\}]=[\left\{ , \right\}] + [S_{\alpha}^0] +
[S_{\alpha}^0].$$ Now let us compose the linear equations. In a
simplest case of the Sclyanin algebra (\cite{Scl}), if we put
$S_\al^{1}$ and $S_0^{1}$ equal to 0, we get the following
equations:

\beq{3p}
\begin{array}{c}
2b + [\{,\}] = a + b,\\
a + b + [\{,\}] = d + 2b\,,
\end{array}
  \eq

where we put  $[S_0] = a$, $[S_\al^{0}] = b$, $[J_\al] = d$. We
get the following important relation $2 [\{,\}] = d$. Note that we
get it without using the value of graduations $a$ and $b$.

Let us consider now the three point fusion. In this case the Lax
operator has the following form:
$$ L_{fusion}(z) =$$
\beq{4p} = \left( S_{0} + S_{0}^{1}E_{2}(z) +
S_{0}^{2}E_{2}^{'}(z) \right) \sigma_{0} + \sum_{\alpha} \left(
S_{\alpha}^{0} \varphi_{\alpha}(z) + S_{\alpha}^{1}
\varphi_{\alpha}^{'}(z) + S_{\alpha}^{2}
\varphi_{\alpha}^{''}(z)\right) \sigma_{\alpha} \,. \eq Analyzing
tables (3) and (4) it is possible to conclude that the "structural
blocks" for the coefficients at $SS$ are the product of the
function $J$. Using the graduation we can write the presumable
form of the exchange relations. Let us consider for example
$\left\{S_{\alpha}^0, S_0\right\}$: \beq{5p} \left\{S_{\alpha}^0,
S_0\right\} = 2i\varepsilon_{\alpha\beta\gamma} \left( J_{\ga \be}
S_0 S_\ga^0 +
 J_{\al} J_{\ga \be} S_\be^1
S_\ga^1 + c_{2,0} S_\be^2 S_\ga^0 + c_{0,2} S_\be^0 S_\ga^2 +
c_{2,2} S_\be^2 S_\ga^2 \right)    \,. \eq Put $[S_0]=0$,
$[S_\al^0]=1$, $[S_0^1]=2$, $[S_\al^1]=2$, $[S_0^2]=3$,
$[S_\al^2]=3$ in accordance with notation above, we get
$[J_{\al}]=-2$ and $[\left\{S_{\alpha}^0, S_0\right\}]=0$. So
there are all allowed terms in (\ref{5p}). For example it is not
possible to write the term with $S_\be^1 S_\ga^2$. Its graduation
is equal to 5, and we cannot pick out the coefficient consisted of
$J$-function product because the graduation of the whole term is
not equal to 0 for each case. Compared the results of the explicit
calculation ((Table 3),(Table 4)), we get the coefficients. The
graduation shows the possible positions of the coefficients. Note
that the coefficients at $SS$ are the invariants of the
transformations which do not change the Poisson brackets.
\begin{center}
Table 3. \ Exchange relations $S_{\alpha},\ S_{\beta}$
\end{center}
{\small
\begin{tabular}{|c|c|c|c|c|c|c|c|c|c|}
\hline \cline{1-1}
$$ & $$ & $$ & $$ & $$ & $$ & $$ & $$ & $$ & $$\\

$2i\varepsilon_{\alpha\beta\gamma} \cdot$ & $S_0S_\ga^{0}$ & $S_0S_\ga^{1}$ & $S_0^{1}S_\ga^{0}$ & $S_0^{1}S_\ga^{1}$ & $S_0S_\ga^{2}$ & $S_0^{2}S_\ga^{0}$& $S_0^{1}S_\ga^{2}$ & $S_0^{2}S_\ga^{1}$ & $S_0^{2}S_\ga^{2}$\\
\hline \cline{1-0}
$$ & $$ & $$ & $$ & $$ & $$ & $$ & $$ & $$ & $$\\

$\left\{S_{\alpha}^0, S_{\beta}^0\right\}$ & $+1$ & $0$ & $-J_\ga$ & $0$ & $0$ & $0$ & $-J_{\ga\be}J_{\ga\al}$ & $J_{\ga\be}J_{\ga\al}$ & $0$\\
\hline \cline{1-0}
$$ & $$ & $$ & $$ & $$ & $$ & $$ & $$ & $$ & $$\\

$\left\{S_{\alpha}^1, S_{\beta}^0\right\}$ & $0$ & $+1$ & $0$ & $-J_\al$ & $0$ & $-J_{\ga\al}$ & $0$ & $0$ & $J_{\ga\al}J_{\al\be}$\\
\hline \cline{1-0}
$$ & $$ & $$ & $$ & $$ & $$ & $$ & $$ & $$ & $$\\

$\left\{S_{\alpha}^0, S_{\beta}^1\right\}$ & $0$ & $+1$ & $0$ & $-J_\be$ & $0$ & $-J_{\ga\be}$ & $0$ & $0$ & $J_{\ga\be}J_{\be\al}$\\
\hline \cline{1-0}
$$ & $$ & $$ & $$ & $$ & $$ & $$ & $$ & $$ & $$\\

$\left\{S_{\alpha}^1, S_{\beta}^1\right\}$ & $0$ & $0$ & $+1$ & $0$ & $+1$ & $0$ & $-(J_\al+J_\be)$ & $0$ & $0$\\
\hline \cline{1-0}
$$ & $$ & $$ & $$ & $$ & $$ & $$ & $$ & $$ & $$\\

$\left\{S_{\alpha}^0, S_{\beta}^2\right\}$ & $0$ & $0$ & $0$ & $0$ & $+1$ & $0$ & $-(J_\ga+J_\be)$ & $J_{\ga\be}$ & $0$\\
\hline \cline{1-0}
$$ & $$ & $$ & $$ & $$ & $$ & $$ & $$ & $$ & $$\\

$\left\{S_{\alpha}^2, S_{\beta}^0\right\}$ & $0$ & $0$ & $0$ & $0$ & $+1$ & $0$ & $-(J_\ga+J_\al)$ & $J_{\ga\al}$ & $0$\\
\hline \cline{1-0}
$$ & $$ & $$ & $$ & $$ & $$ & $$ & $$ & $$ & $$\\

$\left\{S_{\alpha}^1, S_{\beta}^2\right\}$ & $0$ & $0$ & $0$ & $0$ & $0$ & $+1$ & $0$ & $0$ & $-J_{\al\be}$\\
\hline \cline{1-0}
$$ & $$ & $$ & $$ & $$ & $$ & $$ & $$ & $$ & $$\\

$\left\{S_{\alpha}^2, S_{\beta}^1\right\}$ & $0$ & $0$ & $0$ & $0$ & $0$ & $+1$ & $0$ & $0$ & $J_{\al\be}$\\
\hline \cline{1-0}
$$ & $$ & $$ & $$ & $$ & $$ & $$ & $$ & $$ & $$\\

$\left\{S_{\alpha}^2, S_{\beta}^2\right\}$ & $0$ & $0$ & $0$ & $0$ & $0$ & $0$ & $-\frac{1}{2}$ & $+\frac{1}{2}$ & $0$\\
\hline
\end{tabular}
}

\begin{center}
Table 4. \ Exchange relations $S_{\alpha},\ S_{0}$
\end{center}
$${\small
\begin{tabular}{|c|c|c|c|c|c|c|c|c|c|}
\hline
\cline{1-0}
$ $ & $$ & $$ & $$ & $$ & $$ & $$ & $$ & $$ & $$\\
$2i\varepsilon_{\alpha\beta\gamma} \cdot$ & $S_\be^{0} S_\ga^{0}$ & $S_\be^{0}S_\ga^{1}$ & $S_\be^{1}S_\ga^{0}$ & $S_\be^{1}S_\ga^{1}$ & $S_\be^{0}S_\ga^{2}$ & $S_\be^{2} S_\ga^{0}$ & $S_\be^{1}S_\ga^{2}$ & $S_\be^{2}S_\ga^{1}$ & $S_\be^{2}S_\ga^{2}$\\

\hline
\cline{1-0}
$ $ & $$ & $$ & $$ & $$ & $$ & $$ & $$ & $$ & $$\\
$\left\{S_{\alpha}^0, S_{0} \right\}$ & $J_{\ga\be}$ & $0$& $0$ &
$J_\al J_{\ga\be}$ & $-J_{\al}J_{\ga\be}$ & $-J_{\al}J_{\ga\be}$ &
$0$ & $0$ &
$(J_{\be}J_{\ga}-J_{\al}^{2})J_{\ga\be}$\\
\hline
\cline{1-0}
$$ & $$ & $$ & $$ & $$ & $$ & $$ & $$ & $$ & $$\\

$\left\{S_{\alpha}^1, S_0\right\}$ & $0$ & $-J_\be$ & $J_\ga$ & $0$ & $0$ & $0$ & $J_{\ga}J_{\al\be}$ & $J_{\be}J_{\ga\al}$ & $0$\\
\hline \cline{1-0}
$$ & $$ & $$ & $$ & $$ & $$ & $$ & $$ & $$ & $$\\

$\left\{S_{\alpha}^0, S_0^1\right\}$ & $0$ & $0$ & $0$ & $J_{\ga\be}$ & $-J_{\ga\be}$ & $J_{\ga\be}$ & $0$ & $0$ & $(J_{\al\ga}+J_{\al\be})J_{\ga\be}$\\
\hline \cline{1-0}
$$ & $$ & $$ & $$ & $$ & $$ & $$ & $$ & $$ & $$\\

$\left\{S_{\alpha}^1, S_0^1\right\}$ & $0$ & $-1$ & $+1$ & $0$ & $0$ & $0$ & $J_{\al\be}$ & $J_{\ga\al}$ & $0$\\
\hline \cline{1-0}
$$ & $$ & $$ & $$ & $$ & $$ & $$ & $$ & $$ & $$\\

$\left\{S_{\alpha}^2, S_{0}  \right\}$ & $0$ & $0$ & $0$ & $0$ & $-J_{\be}$ & $J_{\ga}$ & $0$ & $0$ & $-J_{\al}J_{\ga\be}$\\
\hline \cline{1-0}
$$ & $$ & $$ & $$ & $$ & $$ & $$ & $$ & $$ & $$\\

$\left\{S_{\alpha}^0, S_{0}^2\right\}$ & $0$ & $0$ & $0$ & $0$ & $0$ & $0$ & $\frac{1}{2}J_{\ga\be}$ & $-\frac{1}{2}J_{\ga\be}$ & $0$\\
\hline \cline{1-0}
$$ & $$ & $$ & $$ & $$ & $$ & $$ & $$ & $$ & $$\\

$\left\{S_{\alpha}^1, S_{0}^2\right\}$ & $0$ & $0$ & $0$ & $0$ & $-1$ & $+1$ & $0$ & $0$ & $J_{\ga\be}$\\
\hline \cline{1-0}
$$ & $$ & $$ & $$ & $$ & $$ & $$ & $$ & $$ & $$\\

$\left\{S_{\alpha}^2, S_{0}^1\right\}$ & $0$ & $0$ & $0$ & $0$ & $-1$ & $+1$ & $0$ & $0$ & $-J_{\ga\be}$\\
\hline\cline{1-0}
$$ & $$ & $$ & $$ & $$ & $$ & $$ & $$ & $$ & $$\\

$\left\{S_{\alpha}^2, S_{0}^2\right\}$ & $0$ & $0$ & $0$ & $0$ & $0$ & $0$ & $-\frac{1}{2}$ & $+\frac{1}{2}$ & $0$\\
\hline
\end{tabular}
}$$
As the result we get the following
\begin{predl}. The space  $\mathcal{P}_{f_{3},2}$ is Poisson with respect to the
corresponding quadratic brackets Table.3 and Table.4.
\end{predl}

\section{Conclusion.}

Although the graduation clarifies the algebraic structure of the
phase space by indicating the possible positions of the
coefficients at $SS$ product but it is not possible to write the
explicit forms of these coefficients. The other question is about
the representation of the coefficients by $J$-function product.
The goal of the further researching can be simulate the structure
of the phase space without using the calculation and abstracting
from the elliptic dependence.

\paragraph{Acknowledgments.}
Author would like to thank A. Levin for fruitful discussions and
M. Olshanetsky for the suggested theme, important discussion and
attention during the writing of article. The work  was partly supported
by grants RFBR-06-02-17381, NSch-8065-2006.2,
RFBR-06-01-92054-KE and by Federal Nuclear Energy Agency.

\section{Appendix.}
\subsection{Appendix A. Elliptic functions}
\setcounter{equation}{0}
\def\theequation{A.\arabic{equation}}

We assume that $q=\exp 2\pi i\tau$, where $\tau$ is the modular
parameter of the elliptic curve $E_\tau$.

The basic element is the theta  function: \beq{A.1a}
\vth(z|\tau)=q^{\frac {1}{8}}\sum_{n\in {\bf Z}}(-1)^n\bfe(\oh
n(n+1)\tau+nz)=~~ (\bfe=\exp 2\pi\imath) \eq

\bigskip

{\it The  Eisenstein functions} \beq{A.1}
E_1(z|\tau)=\p_z\log\vth(z|\tau),
~~E_1(z|\tau)\sim\f1{z}-2\eta_1z, \eq where \beq{A.6}
\eta_1(\tau)=\frac{24}{2\pi i}\frac{\eta'(\tau)}{\eta(\tau)}\,,~~
\eta(\tau)=q^{\frac{1}{24}}\prod_{n>0}(1-q^n)\,. \eq is the
Dedekind function. \beq{A.2} E_2(z|\tau)=-\p_zE_1(z|\tau)=
\p_z^2\log\vth(z|\tau), ~~E_2(z|\tau)\sim\f1{z^2}+2\eta_1\,. \eq

{\it Relation to the Weierstrass functions} \beq{a100}
\zeta(z,\tau)=E_1(z,\tau)+2\eta_1(\tau)z\,,
~~\wp(z,\tau)=E_2(z,\tau)-2\eta_1(\tau)\,. \eq The highest
Eisenstein functions \beq{A.2a}
E_j(z)=\frac{(-1)^j}{(j-1)!}\p^{(j-2)}E_2(z)\,,~~(j>2)\,. \eq

\beq{A.2b} \phi(u,z)\phi(-u,z)=E_2(z)-E_2(u)\,, \eq

\beq{A.3b} \phi(u,z)\phi^{'}(-u,z)= + \left( E_1(u)E_2(u) +
\frac{1}{2}E_2(u)^{'} \right) - E_1(u)E_2(z) +
\frac{1}{2}E_2^{'}(z) \,. \eq

\beq{A.4b} \phi^{'}(u,z)\phi(-u,z)= - \left( E_1(u)E_2(u) +
\frac{1}{2}E_2(u)^{'} \right) + E_1(u)E_2^{'}(z) +
\frac{1}{2}E_2^{'}(z) \,. \eq

$$ \phi^{'}(u,z)\phi^{'}(-u,z)=$$
\beq{A.5b} = \left( - (E_2(u))^2 + (E_1(u))^2 E_2(u) + E_1(u)
E_2^{'}(u) + \frac{1}{3}E_2(u)^{''} \right) - \eq
$$
+ \left( - (E_1(u))^2 + E_2(u) \right)E_2(z) +
\frac{1}{6}E_2^{''}(z) \,.$$

The next important function is \beq{A.3} \phi(u,z)= \frac
{\vth(u+z)\vth'(0)} {\vth(u)\vth(z)}\,. \eq \beq{A.300}
\phi(u,z)=\phi(z,u)\,,~~\phi(-u,-z)=-\phi(u,z)\,. \eq It has a
pole at $z=0$ and \beq{A.3a}
\phi(u,z)=\frac{1}{z}+E_1(u)+\frac{z}{2}(E_1^2(u)-\wp(u))+\ldots\,.
\eq

\beq{A3c} \p_u\phi(u,z)=\phi(u,z) (E_1(u+z)-E_1(u)) \,. \eq

\beq{A3e} \p_z\phi(u,z)=\phi(u,z) (E_1(u+z)-E_1(z)) \,. \eq

\beq{A3d} \lim_{z\to 0}\ln\p_u\phi(u,z)=-E_2(u) \,. \eq

{\it Heat equation} \beq{A.4b} \p_\tau\phi(u,w)-\f1{2\pi
i}\p_u\p_w\phi(u,w)=0\,. \eq

{\it Quasi-periodicity}

\beq{A.11} \vth(z+1)=-\vth(z)\,,~~~\vth(z+\tau)=-q^{-\oh}e^{-2\pi
iz}\vth(z)\,, \eq \beq{A.12}
E_1(z+1)=E_1(z)\,,~~~E_1(z+\tau)=E_1(z)-2\pi i\,, \eq \beq{A.13}
E_2(z+1)=E_2(z)\,,~~~E_2(z+\tau)=E_2(z)\,, \eq \beq{A.14}
\phi(u,z+1)=\phi(u,z)\,,~~~\phi(u,z+\tau)=e^{-2\pi \imath
u}\phi(u,z)\,. \eq \beq{A.15}
\p_u\phi(u,z+1)=\p_u\phi(u,z)\,,~~~\p_u\phi(u,z+\tau)=e^{-2\pi
\imath u}\p_u\phi(u,z)-2\pi\imath\phi(u,z)\,. \eq

 {\it  The Fay three-section formula:}
\beq{ad3}
\phi(u_1,z_1)\phi(u_2,z_2)-\phi(u_1+u_2,z_1)\phi(u_2,z_2-z_1)-
\phi(u_1+u_2,z_2)\phi(u_1,z_1-z_2)=0\,. \eq

{\it  From (\ref{A3e}) and (\ref{ad3}) we have:}

\beq{ad31} \phi(u_1,z)\phi(u_2,z)=\phi(u_1+u_2,z)(E_1(u_1)+
E_1(u_2) - E_1(u_1 + u_2 + z) + E_1(z)) \,. \eq

Particular cases of this formula are the  functional equations
\beq{ad2}
\phi(u,z)\p_v\phi(v,z)-\phi(v,z)\p_u\phi(u,z)=(E_2(v)-E_2(u))\phi(u+v,z)\,,
\eq \beq{i}
\phi(u,z_1)\phi(-u,z_2)=\phi(u,z_1-z_2)(-E_1(z_1)+E_1(z_2)-E_{1}(u)+E_{1}(u+z_{1}-z_{2}))=
\eq
$$
 = \phi(u,z_1-z_2)(-E_1(z_1)+E_1(z_2)+\p_u\phi(u,z_2-z_1))\,,
$$
\beq{ir1} \phi(u,z)\phi(-u,z)=E_2(z)-E_2(u)\,.
 \eq

\beq{ir} \phi(v,z-w)\phi(u_1-v,z)\phi(u_2+v,w)
-\phi(u_1-u_2-v,z-w)\phi(u_2+v,z)\phi(u_1-v,w)= \eq
$$
\phi(u_1,z)\phi(u_2,w)f(u_1,u_2,v)\,,
$$
where \beq{ir8} {\bf
f}(u_1,u_2,v)=E_1(v)-E_1(u_1-u_2-v)+E_1(u_1-v)-E_1(u_2+v)\,. \eq
One can rewrite the last function as \beq{ir3} {\bf
f}(u_1,u_2,v)=-\frac{ \vth'(0)\vth(u_1)\vth(u_2)\vth(u_2-u_1+2v)
}{ \vth(u_1-v)\vth(u_2+v)\vth(u_2-u_1+v)\vth(v) }\,. \eq

Using (\ref{A.1}), (\ref{A.2}), (\ref{A.3a}) one can derive from
(\ref{ir}) some important particular cases. One of them
corresponding to $v=u_1$ (or $v=-u_2$), is the Fay identity
(\ref{ad3}). Another particular case comes from $u_1=0$ (or
$u_2=u$): \beq{ir7}
\phi(v,z-w)\phi(-v,z)\phi(u+v,w)-\phi(-u-v,z-w)\phi(u+v,z)\phi(-v,w)=
\eq
$$
\phi(u_1,z)(E_2(u+v)-E_2(v))\,.
$$
If $u_2\to -v$ then (\ref{ir}) in the first non-trivial order take
the form for $u_1=\al,~u_2=\be$ \beq{ir4}
\phi(-\be,z-w)E_1(w)\phi(\al+\be,z)-
\phi(\al,z-w)E_1(z)\phi(\al+\be,w)= \eq
$$
\phi(\al,z)\phi(\be,w)(E_1(\al)+E_1(\be)-E_1(\al+\be))\,.
$$

\subsection{Appendix B.  Lie algebra $\sln$, Group $GL(\mathcal{N},\mathbb{C})$
 and elliptic functions}
\setcounter{equation}{0}
\def\theequation{B.\arabic{equation}}

Introduce the notation
$$
{\bf e}_N(z)=\exp (\frac{2\pi i}{N} z)
$$
 and two matrices
\beq{q} Q=\di({\bf e}_N(1),\ldots,{\bf e}_N(m),\ldots,1) \eq
\beq{la} \La=\de_{j,j+1}\,,~~(j=1,\ldots,N\,,~mod\,N)\,.
 \eq
 Let
\beq{B.10}
\mZ^{(2)}_N=(\mZ/N\mZ\oplus\mZ/N\mZ)\,,~~\ti{\mZ}^{(2)}_N)=
\mZ^{(2)}_N\setminus(0,0) \eq be the two-dimensional lattice of
order $N^2$ and $N^2-1$ correspondingly. The matrices
$Q^{a_1}\La^{a_2}$, $a=(a_1,a_2)\in\mZ^{(2)}_N$ generate a basis
in the group $\GLN$, while $Q^{\al_1}\La^{\al_2}$,
$\al=(\al_1,\al_2)\in\ti{\mZ}^{(2)}_N$ generate a basis in the Lie
algebra $\sln$. More exactly, we introduce the following basis in
$\GLN$. Consider the projective representation of $\mZ^{(2)}_N$ in
$\GLN$ \beq{B.11} a\to T_{a}= \frac{N}{2\pi
i}\bfe_N(\frac{a_1a_2}{2})Q^{a_1}\La^{a_2}\,, \eq \beq{AA3a}
T_aT_b=\frac{N}{2\pi i}\bfe_N(-\frac{a\times b}{2})T_{a+b}\,, ~~
(a\times b=a_1b_2-a_2b_1) \,. \eq Here $\frac{N}{2\pi i}
\bfe_N(-\frac{a\times b}{2})$ is a non-trivial two-cocycle in
$H^2(\mZ^{(2)}_N,\mZ_{2N})$. The matrices $T_\al$,
$\al\in\ti{\mZ}^{(2)}_N$ generate a basis in $\sln$. It follows
from (\ref{AA3a}) that \beq{AA3b}
[T_{\al},T_{\be}]=\bfC(\al,\be)T_{\al+\be}\,, \eq where
$\bfC(\al,\be)=\frac{N}{\pi}\sin\frac{\pi}{N}(\al\times \be)$ are
 the structure constants of $\sln$.

 For $N=2$ the basis $T_{\al}$ is proportional to the basis of the Pauli
matrices:
 $$
 T_{(1,0)}=\f1{\pi\imath}\si_3\,,~~
 T_{(0,1)}=\f1{\pi\imath}\si_1\,,~~
  T_{(1,1)}=\f1{\pi\imath}\si_2\,.
  $$

 The Lie coalgebra $\gg^*=\sln$ has the dual basis
  \beq{db}
 \gg^*=\{\bfS=\sum_{\ti{\mZ}^{(2)}_N}S_\ga t^\ga\}\,,~~
 t^\ga=\frac{2\pi\imath}{N^2}T_{-\ga}\,,~~\lan T_\al
t^\be\ran=\de_{\al}^{-\be}\,.
 \eq
It follows from (\ref{AA3b}) that $\gg^*$ is a Poisson space with
the linear brackets \beq{A101}
\{S_\al,S_\be\}=\bfC(\al,\be)S_{\al+\be}\,. \eq The coadjoint
action in these basises   takes the form \beq{coad} {\rm
ad}^*_{T_\al}t^\be=\bfC(\al,\be)t^{\al+\be}\,. \eq

Let $\bfga=\frac{\ga_1+\ga_2\tau}{N}$. Then introduce the
following  constants on $\ti{\mZ}^{(2)}$: \beq{AA50}
\vth(\bfga)=\vth\bigl(\frac{\ga_1+\ga_2\tau}{N}\bigr)\,, ~~
E_1(\bfga)=E_1\bigl(\frac{\ga_1+\ga_2\tau}{N}\bigr)\,,
~~E_2(\bfga)=E_2\bigl(\frac{\ga_1+\ga_2\tau}{N}\bigr)\,, \eq
\beq{ph} \phi_\ga(z)=\phi(\bfga,z)\,, \eq \beq{vf}
\vf_\ga(z)=\bfe_N(\ga_2z)\phi_\ga(z)\,, \eq \beq{vf1}
\vf_{\gamma,\eta}(z)=\bfe_N(\ga_2z)\phi(\eta +
\frac{\ga_1+\ga_2\tau}{N},z)\,. \eq They have the following
quasi-periodicities \beq{qpe1}
\vf_\ga(z+1)=\bfe_N(\ga_2)\vf_\ga(z)\,,~~
\vf_\ga(z+\tau)=\bfe_N(-\ga_1)\vf_\ga(z)\,, \eq \beq{qpe2}
\vf_{\ga,\eta}(z+1)=\bfe_N(\ga_2)\vf_{\ga,\eta}(z)\,,~~
\vf_{\ga,\eta}(z+\tau)=\bfe_N(-\ga_1-\eta)\vf_{\ga,\eta}(z)\,, \eq

The important formulas with $\vf_\al(z)$

\beq{vf1} \vf_\ga^{'}(z) = -\vf_\al(z)\vf_\be(z) =
\vf_\ga(z)(E_{1}(\ga+z)-E_{1}(z)-E_{1}(\ga)) \,, \eq

\beq{vf2} \vf_\ga^{''}(z) = \vf_\ga(z)(E_{1}^{'}(\al) +
E_{1}^{'}(\be)) - 2\vf_\ga(z)E_{1}^{'}(z) \,. \eq



\newpage
\begin{thebibliography}{0}

\bibitem{Ch}\ Yu. Chernyakov, {\it TMF}, {\bf T 141} (2004) 1, 38-59;

\bibitem{MPR}\ Musso F., Petrera M., Ragnisco O.,
{\it Jour. Nonlinear Math. Phys.}, {\bf 12 suppl.} 1, (2005)
482-498;

\bibitem{IW}\
E.Inonu, E.Wigner, On contraction of groups and their
representations, {\it Proc. Nat. Acad. Sci.}, {\bf 39} (1953),
510-24;

\bibitem{ChLOZ}\ Yu.Chernyakov, A.M.Levin,
M.Olshanetsky, A.Zotov, {\it Jour. of Phys.A}, {\bf v39}, 39,
(2006) 12083-12101;

\bibitem{Sch}\ Schlesinger L., {\it J.Reine Angew.Math}, {\bf 141}, (1912) 96-145;

\bibitem{LOZ}\
A.Levin, M.Olshanetsky, A.Zotov, Painleve VI, Rigid Tops and
Reflection Equation,
 math.QA/0508058,   submitt. to {\it Comm. Math. Phys.};

\bibitem{Ta}\
K.Takasaki, {\it Lett.Math.Phys.}, {\bf 44}, (1998) 143--156.
hep-th/9711058;

\bibitem{LO}\
 A.Levin, M.Olshanetsky,
  {\it Amer. Math. Soc.
Transl. Ser. 2}, {\bf 191},
 Amer. Math. Soc., Providence, RI, (1999), 223--262,
 hep-th/9709207;

\bibitem{Scl}\ E.Sklyanin, {\it Func. Anal. Appl.}
{\bf Vol.16}, (1982) 283.

\end{thebibliography}
\end{document}